\documentclass[conference]{IEEEtran}
\IEEEoverridecommandlockouts

\usepackage{cite}
\usepackage{amsmath,amssymb,amsfonts}
\usepackage{algorithmic}
\usepackage{graphicx}
\usepackage{textcomp}
\usepackage{xcolor}
\usepackage{url}
\usepackage{threeparttable}
\def\BibTeX{{\rm B\kern-.05em{\sc i\kern-.025em b}\kern-.08em
    T\kern-.1667em\lower.7ex\hbox{E}\kern-.125emX}}
\usepackage{caption}
\usepackage{blindtext}
\usepackage{tcolorbox}
\usepackage[final]{pdfpages}
\usepackage{lipsum,multicol}
\usepackage{xcolor}
\usepackage{tikz}
\usepackage{listings}
\usepackage{enumitem}
\usepackage{hyperref}
\usepackage{amsfonts}
\usepackage{wrapfig}
\usepackage{subcaption} 
\usepackage{adjustbox}
\usepackage{colortbl}
\usepackage{fancybox}
\usepackage{multirow}
\usepackage[normalem]{ulem}
\useunder{\uline}{\ul}{}
\usepackage{enumitem}
\usepackage{booktabs}

\definecolor{main}{HTML}{5989cf}    
\definecolor{sub}{HTML}{cde4ff}     

\newtcolorbox{boxB}{
    fontupper = \bf\color{main}\footnotesize, 
    boxrule = 0.5pt,
    colframe = main,
    rounded corners,
    arc = 5pt   
}

\newtcolorbox{boxD}{
    fontupper = \small, 
    colback = sub, 
    colframe = main, 
    boxrule = 0pt, 
    toprule = 2pt, 
    bottomrule = 2pt 
}

\newtcolorbox{boxH}{
    fontupper = \small, 
    colback = sub, 
    colframe = main, 
    boxrule = 0pt, 
    leftrule = 6pt 
}

\newtcolorbox{boxG}{
    enhanced,
    boxrule = 0pt,
    colback = sub,
    borderline west = {1pt}{0pt}{main}, 
    borderline west = {0.75pt}{2pt}{main}, 
    borderline east = {1pt}{0pt}{main}, 
    borderline east = {0.75pt}{2pt}{main}
}    

\newtcolorbox{boxK}{
    fontupper = \small,
    sharpish corners, 
    boxrule = 0pt,
    toprule = 1.0pt, 
    enhanced,
    fuzzy shadow = {0pt}{-2pt}{-0.5pt}{0.5pt}{black!35} 
}

\definecolor{MidnightBlue}{HTML}{006895}

\newcommand*\circled[1]{\tikz[baseline=(char.base)]{
            \node[shape=circle,draw,inner sep=0.5pt] (char) {#1};}}

\newboolean{showcomments}

\setboolean{showcomments}{true}

\ifthenelse{\boolean{showcomments}}
  {\newcommand{\nb}[2]{
    \fbox{\bfseries\sffamily\scriptsize#1}
    {\sf\small$\blacktriangleright$\textit{#2}$\blacktriangleleft$}
   }
   
  }
  {\newcommand{\nb}[2]{}
   
  }

\newcommand{\ie}{\textit{i.e.,}\xspace}
\newcommand{\eg}{\textit{e.g.,}\xspace}

\newcommand{\etal}{et al.\xspace}

\newcommand{\llms}{LLMs\xspace}
\newcommand{\llm}{LLM\xspace}

\newcommand{\github}{GitHub\xspace}

\newcommand{\randomCut}{\textit{RandomCut}\xspace}

\newcommand{\commitGen}{\textit{FromCommit}\xspace}
\newcommand{\summarizationGen}{\textit{SummarizationGen}\xspace}

\newcommand{\ASTerrors}{\textit{n\_ast\_errors}\xspace}
\newcommand{\ASTlevels}{\textit{n\_ast\_levels}\xspace}
\newcommand{\ASTnodes}{\textit{n\_ast\_nodes}\xspace}
\newcommand{\whitespaces}{\textit{n\_whitespaces}\xspace}
\newcommand{\complexity}{\textit{complexity}\xspace}
\newcommand{\nloc}{\textit{nloc}\xspace}
\newcommand{\tokenCount}{\textit{token\_count}\xspace}
\newcommand{\identifiers}{\textit{n\_identifiers}\xspace}
\newcommand{\commitID}{\textit{commit\_id}\xspace}
\newcommand{\funName}{\textit{fun\_name}\xspace}
\newcommand{\commitMessage}{\textit{commit\_message}\xspace}
\newcommand{\docstring}{\textit{docstring}\xspace}

\newcommand{\snipgen}{\textit{SnipGen}\xspace}

\newcommand{\secref}[1]{Sec.~\ref{#1}\xspace}
\newcommand{\figref}[1]{Fig.~\ref{#1}\xspace}
\newcommand{\tabref}[1]{Table~\ref{#1}\xspace}

\begin{document}
\bstctlcite{IEEEexample:BSTcontrol}
\title{
SnipGen: A Mining Repository Framework for Evaluating LLMs for Code 
}

\author{\IEEEauthorblockN{
Daniel Rodriguez-Cardenas, Alejandro Velasco, and
Denys Poshyvanyk}
\IEEEauthorblockA{Department of Computer Science,
William \& Mary\\
Williamsburg, VA\\
Email: dhrodriguezcar, svelascodimate, dposhyvanyk\{@wm.edu\}}}


\maketitle

\begin{abstract}
Large Language Models (\llms), such as transformer-based neural networks trained on billions of parameters, have become increasingly prevalent in software engineering (SE). These models, trained on extensive datasets that include code repositories, exhibit remarkable capabilities for SE tasks. However, evaluating their effectiveness poses significant challenges, primarily due to the potential overlap between the datasets used for training and those employed for evaluation. To address this issue, we introduce \snipgen, a comprehensive repository mining framework designed to leverage prompt engineering across various downstream tasks for code generation. \snipgen aims to mitigate data contamination by generating robust testbeds and crafting tailored data points to assist researchers and practitioners in evaluating \llms for code-related tasks. In our exploratory study, \snipgen mined approximately $227K$ data points from $338K$ recent code changes in GitHub commits, focusing on method-level granularity. \snipgen features a collection of prompt templates that can be combined to create a Chain-of-Thought-like sequence of prompts, enabling a nuanced assessment of \llms' code generation quality. By providing the mining tool, the methodology, and the dataset, \snipgen empowers researchers and practitioners to rigorously evaluate and interpret \llms' performance in software engineering contexts.

\end{abstract}


\section{Introduction}\label{sec:introduction}

Large Language Models (\llms) have demonstrated significant success across diverse software engineering (SE) tasks, including code auto-completion \cite{austin2021program, Hendrycks2021apps, chen_generation_2021,White.MSR.2015,Ciniselli.TSE}, code summarization \cite{leclair_ensemble_2021,Moran.SANER.2022}, code review \cite{Tufano.ICSE.2021, Tufano.ICSE.2022}, code translation \cite{Nguyen:ICSE15},  clone detection \cite{White.ASE2016,Tufano.MSR.2018}, and program repair \cite{Tufano2019LearningBugFixes,zhou_devign_nodate,SANER.2019,Tufano.ICSE19.Changes,ASE.2018,Zimin.Sequencer,CanWeFix}. \llms are neural models trained on huge datasets including complete \github repositories. Common testbeds for evaluating LLMs for code such as \texttt{HumanEval}, \texttt{MBPP} and \texttt{CodeXGlue} are no longer sufficient \cite{jain_livecodebench_2024}. In addition, as benchmarks and testbeds are released, new \llms probably already seen those testbeds. Therefore the testbeds are prone to be outdated as soon as a new LLM is released.



\llms perform complex tasks by relying on statistical knowledge acquired from data distributions, a phenomenon described by Wei \etal as \textit{emerging capabilities} \cite{wei_emergent_2022}.  Given the limited understanding of the nature of this phenomenon, we can formulate an important question: \textit{under what conditions \llms produce the desired output?} Prompt engineering addresses this question by harnessing these capabilities, guiding \llms to make more accurate predictions. Furthermore, given that \llms can extract rules from the provided context (\ie in-context learning), prompt engineering is a natural and intuitive way for people to use \llms.


Recent studies have demonstrated that \llms exhibit improvements in accuracy for downstream tasks when prompts are enhanced and augmented \cite{Zhou2022LargeLM, White2023ChatGPTPP}. Moreover, new methods for crafting better prompts are being explored. For example, Beurer-Kellner \etal \cite{promptingIsProgramming} introduce the idea of Language Model Programming (LMP) which combines text-based prompting with scripting. Furthermore, Wei \etal \cite{wei_chain--thought_2023} shows that the incorporation of Chain-of-Thought (CoT) significantly improves the ability of \llms to perform complex reasoning.



Understanding the internal mechanisms of \llms presents a significant challenge. Current datasets and benchmarks often lack the curated data necessary for thorough performance analyses. Therefore, there is a critical need for consistent data points to effectively evaluate the performance of \llms across various SE tasks. We argue that well-designed testbeds and prompts are the key to accurately assessing \llms understanding of complex information, such as task-related semantics.


To bridge the gap between existing datasets and benchmarks, we developed \snipgen. \snipgen is a framework to collect source code snippets from \github. Each snippet is automatically augmented with prompts tailored for various software tasks. Practitioners and researchers can query and generate new prompts according to the SE task and experiment with different configurations for evaluating \llms for code.  Our goal is to provide resources that can more accurately assess the performance of \llms and aid in the construction of more detailed benchmarks.

The contributions of this paper are listed as follows: 1) A Framework for mining software repositories and crafting data points augmented with prompts for specific SE downstream tasks. 2) a generated testbed comprising Python snippets with calculated features from the AST, natural language, and vulnerabilities analysis~\cite{zenodo_14279563}. 3) Prompt-generated dataset with mutated snippets crafted for Code Completion, Commit generation, and Code summarization. 4) source code and complementary material used in this research are published in an open-source repository\cite{snipgen}.
\section{The \snipgen Framework}\label{sec:methodology}

\snipgen is a framework to extract snippets at method granularity from \github. \snipgen follow steps for curating the extracted raw data and take features from the data such as the number of identifiers, vocabulary, tokens, etc. Features derived from their AST representations and further complementary data. Our dataset can potentially improve the quality of the predictions in downstream tasks by augmenting the prompts, thereby enabling \llms to perform more effectively.

\begin{figure}[ht]
		\centering
		\includegraphics[width=0.48\textwidth]{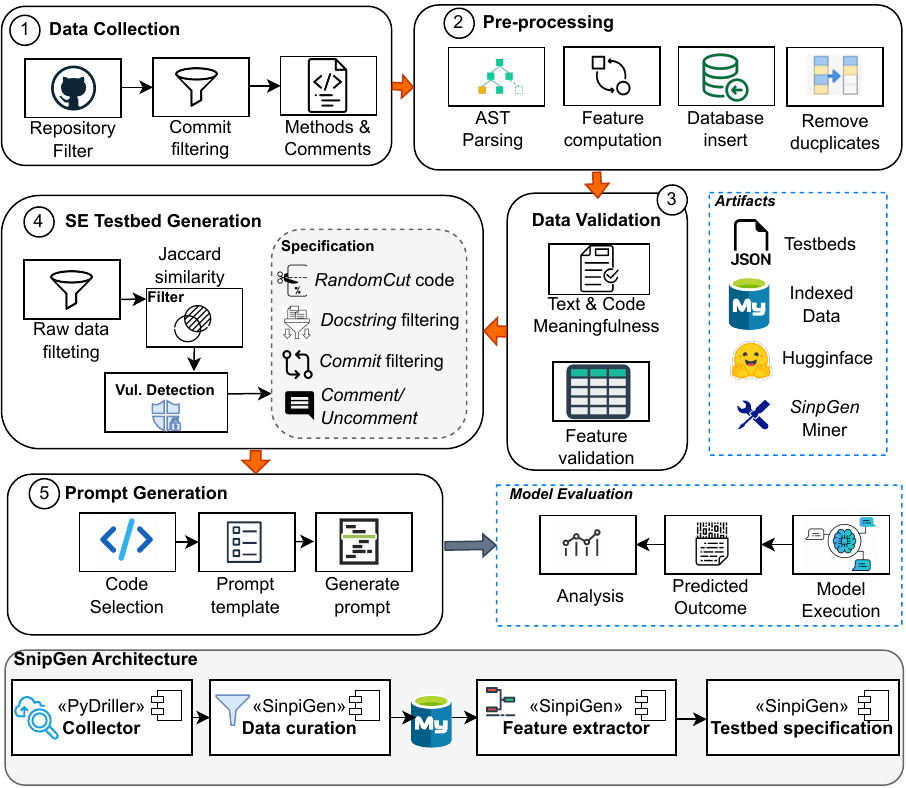}
		\caption{\snipgen Data collection and prompt generation }
        \label{fig:collection}
\end{figure}

Fig. \ref{fig:collection} depicts the process followed by \snipgen to generate a testbed and the \snipgen architecture. The \snipgen architecture comprises components to collect, curate, store, extract, and generate a SE-oriented testbed. The process begins with \circled{1}, the Data Collection phase, where source code snippets are extracted with \textit{pydriller} library~\cite{pydriller} from selected repositories in \github given a tie window. A set of snippets representing a Python method is extracted from each commit. This is followed by \circled{2}, a Pre-processing step, where the -Data Curation- \snipgen component stores the raw data in a \textit{MySQL} database. Once the data is formatted and saved in the storage, \snipgen looks for exact match snippets and removes duplicates. The -Feature extractor- component parses the code into the AST representation using tree-sitter~\cite{tree_sitter} and computes associated features (\ie the number of AST levels, AST nodes, comments, function name).

The data validation at step \circled{3} is a manual evaluation where the authors confirm the dimension values and the meaningfulness of the \textit{Docstring} and linked code, the two authors first selected the docstring with more than 20 words and evaluate the description against the code snippet. The description must depict the steps or intention of the snippet.

The testbed generation step \circled{4}, filters the raw data, evaluates the Jaccard similarity, and identifies vulnerable code. The raw filtering depends on the SE task, for example for code completion \snipgen filters the valid code with more than two lines of code. \snipgen uses CodeQL~\cite{codeql_overview} for vulnerability detection and appends the vulnerability location on the snippet. Finally, step \circled{5} uses the selected snippets from \circled{4} and applies the prompt template to the aimed SE task generating a final prompt. 
\snipgen enables the model evaluation and benchmarking as used in \cite{galeras, astexplainer, syntax_capabilities}.
The following subsections include a detailed description of the features of each data point.

\subsection{Data Point Feature Structure}
\begin{figure}[h]
		\centering
		\includegraphics[width=0.48\textwidth]{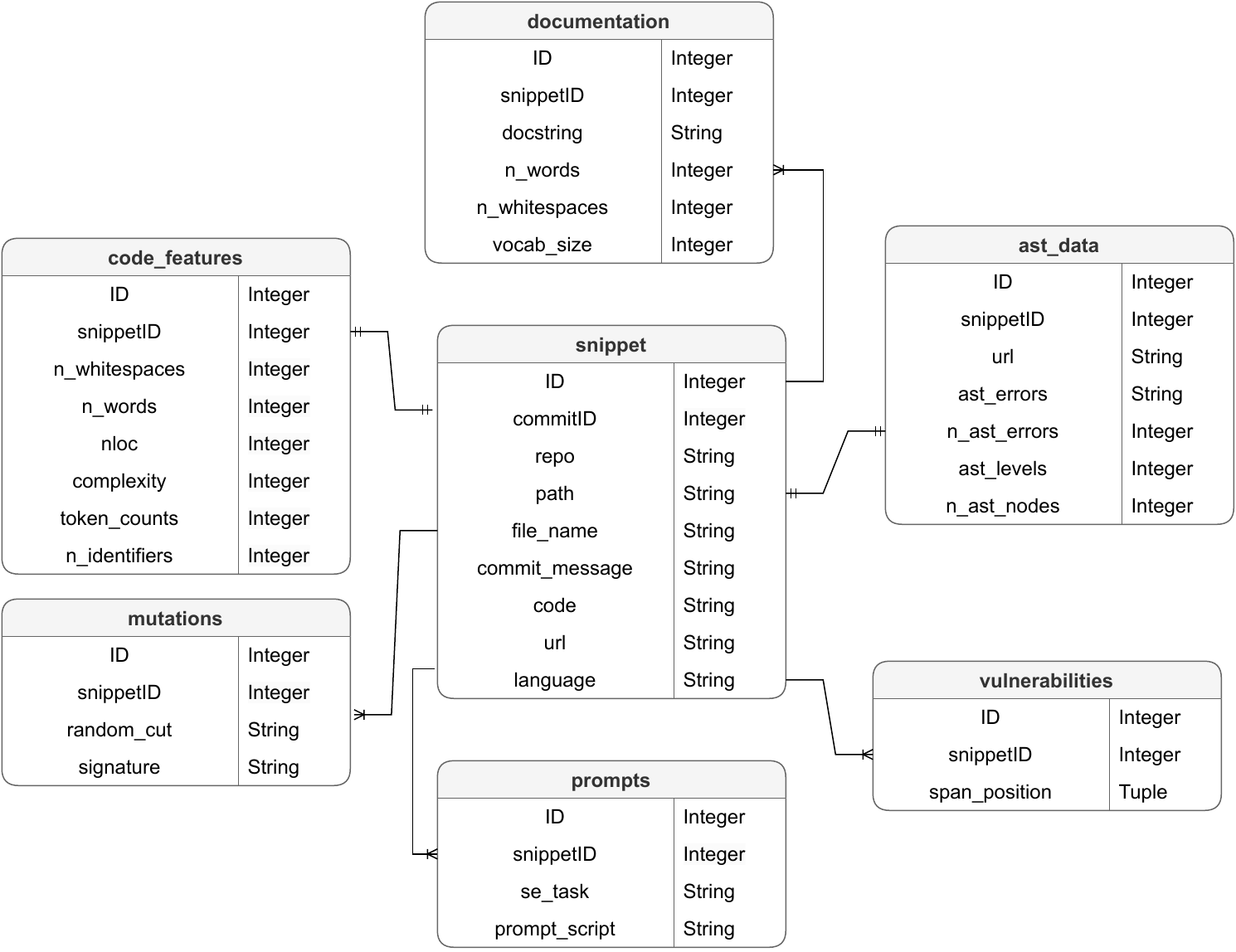}
		\caption{\snipgen data schema. The snippet represents the core commit collected with documentation.  Linked tables contain calculated features. }
        \label{fig:diagram}
\end{figure}
\snipgen can collect a set of Python methods that serve as evaluative data points. Each data point has associated features at seven dimensions as observed at \figref{fig:diagram}. These seven dimensions describe the static feature from the snippet. We aim to link code fragments with their properties. The first dimension corresponds to snippets' identification,  which includes the \commitID (\ie commit hash), \textit{repository} name, \textit{path}, \textit{file\_name}, \funName, \commitMessage. The second dimension is related to the associated documentation \docstring. The \docstring extended to complementary natural language features such as \textit{n\_words, vocab\_size, language,} and \whitespaces. The third dimension corresponds to the snippet's syntactic information, which includes the actual code base, \ASTerrors, \ASTlevels, \ASTnodes,\textit{n\_words}, \textit{vocab\_size}, \tokenCount, and \whitespaces. The fourth dimension corresponds to canonical software metrics, which include \nloc, \complexity, \identifiers. The fifth dimension depicts the span position for vulnerabilities detected from the \textit{code} snippet. The sixth dimension is associated with the snippet mutation when the code is randomly cut one line after the signature, therefore \snipgen identifies the signature as the original \textit{snippetID} and cut code. Finally, the seventh dimension comprises the linked features to the generated prompt. \snipgen labels the prompt to the SE task and the prompt configuration.

\subsection{Software Engineering Tesbed Task}
\label{sec:se_tasks}

Data curation, pre-processing, and data validation produce a testbed oriented to evaluate a model. For instance, a \textit{RandomCut} and \textit{WithDocString} testbeds might evaluate the model at SE tasks, such as \textit{\textbf{code completion}}—generating code to fill in missing parts of a function. \textit{WithDocString} testbed selects the snippets with valid documentation and code so the \llm input compresses both a description and code.
\textit{FromCommit} testbed is focused on selecting meaningful commit messages and linked source code so that to evaluate either \textit{\textbf{commit generation}}—producing commit messages based on code changes or \textit{\textbf{code generation}} producing the complete snippet from the description.  \textit{FromDocString} testbed select only meaningful code descriptions (\ie only \docstring) to generate the code snippet also configuring a code generation case.
\snipgen can be used to evaluate \textit{\textbf{code summarization}}—creating natural language descriptions of the functionality implemented in the provided source code. If we select the original code from the \textit{WithDocString} testbed and the ones at the top of docstring length then we can use the testbed for summarization.

\subsection{Prompt templates}
\label{sec:prompt_templates}

\begin{table}[t]
\centering
\caption{Prompt templates for each SE task using collected \snipgen features.}
\label{tab:templates}

\scalebox{0.7}{%
\setlength{\tabcolsep}{5pt} 
\begin{tabular}{lllp{3.6in}}
\toprule
\multicolumn{1}{c}{\textbf{SE Task}} &
  \multicolumn{1}{c}{\textbf{ID}} &
   &
  \multicolumn{1}{c}{\textbf{Prompt Template}} \\ \hline 
\multirow{3}{*}{\textit{\textbf{\begin{tabular}[c]{@{}l@{}}Code \\ completion\end{tabular}}}} &
  $P1$ &
   &
  Complete the following \textless{}\textit{language}\textgreater method: \textless{}\randomCut\textgreater{} \\
 &
  $P2$ &
   &
  You have a \textless{}\textit{language}\textgreater function named \textless{}\textit{signature}\textgreater{}, the function starts with the following code \textless{}\randomCut\textgreater{}. The function is in charge of \textless{}\docstring\textgreater{} \\
 &
  $P3$ &
   &
  Create a function that accomplish the following functionality in \textless{}\textit{language}\textgreater code: \textless{}\docstring\textgreater{} \\ \hline 
\textit{\textbf{\begin{tabular}[c]{@{}l@{}}Commit \\ generation\end{tabular}}} &
  $P4$ &
   &
  Please describe the following code change to create log message: status before \textless{}\randomCut\textgreater status now \textless{}code\textgreater{} \\ \hline 
\textit{\textbf{Summ.}} &
  $P5$ &
   &
  I need a summary for the following code: \textless{}\textit{code}\textgreater{} \\ \hline 
\multirow{3}{*}{\textit{\textbf{\begin{tabular}[c]{@{}l@{}}Processing \\ prompt\end{tabular}}}} &
  $P6$ &
   &
  Change the method signature by \textless{}signature\textgreater{} \\
 &
  $P7$ &
   &
  Reduce or complete the method using only \textless{}\nloc\textgreater lines of code \\
 &
  $P8$ &
   &
  remove comments; remove summary; remove throws; remove function modifiers\\
  \bottomrule
\end{tabular}
\vspace{-0.3cm}
}
\end{table}

The effectiveness of \llms in code generation is greatly influenced by prompt design. At this point \snipgen only produces a set of data points that can be organized as an input for an autoregressive \llm since the tesbed contains the input and the expected output. \snipgen combines prompt templates and gathered data points to build the final prompt input. The structure, keywords, and context of a prompt play a crucial role in shaping results and analyses. Prompts can be configured as a single-step or multi-step, with the latter allowing iterative refinement based on the model's initial response. Chau \etal \cite{liu_improving_2023} explore such multi-step configurations. Table \ref{tab:templates} lists eight prompt templates, practitioners can modify the template according to the evaluation task. From the proposed list, $P1-P5$ supports single-step SE tasks, while $P6-P8$ enables multi-step processing by combining prompts to refine outputs. For instance, \snipgen can combine $P1+P8$, $P3+P6$, or $P3+P8$ for code completion.


For \textit{\textbf{code completion}}, \snipgen defines three prompts. $P1$ asks the model to complete a method from a randomly selected cut position (\randomCut) in the specified programming language (\textit{language}). $P2$ extends $P1$ by including additional details, such as the method's \textit{signature} and \textit{docstring}. $P3$, in contrast, provides only an NL description extracted from the method's \textit{docstring}. In commit generation, $P4$ instructs the model to create an NL description of the changes made to transform the \randomCut version of a method into its complete code (\textit{code}). $P5$ is designed to ask the model to generate the commit message from the mutated code and the actual code. Lastly, in \textit{\textbf{code summarization}}, \textit{P6} provides only the code, which the model uses to generate a corresponding summary.


%




\subsection{\snipgen Prompt Generation and Use}

\begin{figure}[ht]
		\centering
		\includegraphics[width=0.5\textwidth]{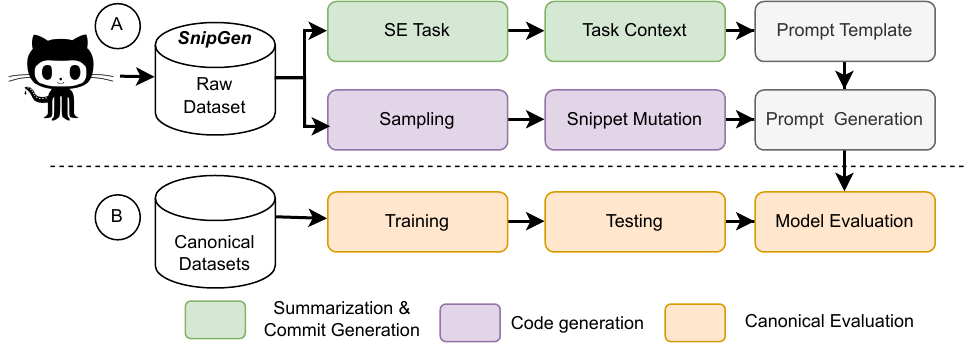}
		\caption{\snipgen Dataset and use. \textit{A} describes the \snipgen data collection and steps until prompt generation. \textit{B} describes the canonical path for training and evaluate \llms}
        \label{fig:overview}
\end{figure}

The \snipgen framework is designed to select a SE task and evaluate a \llm using the testbed with a given context with a designed prompt. \figref{fig:overview} depicts the options a practitioner has to evaluate a \llm. The database supports a query to filter the snippets according to the SE task, for instance for code completion we can sample the snippets with linked \docstring with more than 10 words, this provides the task context (see, Fig. \ref{fig:overview} section \circled{A}). The prompt generation might contain a mutated snippet such as \randomCut to perform the required task. For example, for code completion, we will need a partial code snippet that must be auto-completed by the \llm therefore we need to cut the original snippet smartly. \snipgen can use the \randomCut method to split the code beyond the method signature. Practitioners can still evaluate the model using \textit{canonical} datasets and metrics to compare against the new collected \snipgen testbed.

\section{Experience Report}\label{sec:experience}

In this section, we describe our experience of using \snipgen for collecting a testbed and generating a set of prompts. We also briefly describe three use cases illustrating how \snipgen was successfully used to evaluate \llms for code.

\subsection{\snipgen Testbed Generation}

The experience with \snipgen begins by mining repositories from \github, as detailed in \secref{sec:methodology}. We focused on the most popular Python repositories, applying the following query filters: \textit{language:Python fork:false size:$>=30000$ pushed:$>$2021-12-31 stars: $> 2000$.} The query gathers the most popular repositories in Python. We selected the top 200 repositories including \textit{keras, numpy, pandas, sentry, etc.} We extracted the new snippets reported on commits between 2022 and 2023 from selected repositories. Then we used the \textit{data curation} to remove duplicates and \textit{feature extraction} to generate and extract the associated features. We configured a $0.7$ similarity threshold~\cite{Allamanis19, wang_neural_2019} to de-duplicate snippets using HuggingFace tokenizer BPE. \snipgen saves the raw data and their features into a JSON and a database.  We randomly validated $960$ out of $\approx 227K$  data points to confirm the extracted features and the meaningfulness of the \textit{Docstring} and linked code.

We sampled until $5k$ data points from the \textit{RawData} testbed to construct six testbeds, each tailored for a specific SE task as described at \secref{sec:se_tasks}. To create \randomCut, we selected data points with more than $10$ tokens or $100$ characters, and subsequently, each data point was randomly truncated after the method signature. For \summarizationGen and \commitGen, we filtered \textit{RawDataDocstring} data points with more than 10 words or 50 characters. Table.~\ref{tab:dedupe} provides information about the SE task associated with each curated testbed, the percentage of detected duplicates, the final size, and the generated number of prompts.

\begin{table}[t]
\centering
\caption{Dataset size and deduplication percentage}
\label{tab:dedupe}

\scalebox{0.72}{%
\setlength{\tabcolsep}{5pt} 

\begin{tabular}{llccccc}
\toprule
\multicolumn{1}{c}{\textbf{SE Task}} &
  \multicolumn{1}{c}{\textbf{Testbed}} &
  \textbf{I/O} &
  \textbf{Dupes} &
  \textbf{Dupe \%} &
  \textbf{Size} &
  \textbf{Prompts} \\ \hline
\multirow{2}{*}{\textbf{\begin{tabular}[c]{@{}l@{}}Code \\ Completion\end{tabular}}} &
  \textit{RandomCut} &
  code $\Rightarrow$ code &
  120 &
  2.4\% &
  4880 &
  9760 \\
                         & \textit{WithDocString}     & code\&text $\Rightarrow$ code & 145 & 2.9\% & 4855 & 9710 \\ \hline
\multirow{2}{*}{\textbf{\begin{tabular}[c]{@{}l@{}}Code \\ Generation\end{tabular}}} &
  \textit{FromDocString} &
  text $\Rightarrow$ code &
  76 &
  1.5\% &
  4924 &
  14772 \\
                         & \textit{FromCommit}        & text $\Rightarrow$ code       & 97  & 1.9\% & 4903 & 4903 \\ \hline
\textbf{Sumarization}    & \textit{SummarizationGen}  & code $\Rightarrow$ text       & 156 & 3.1\% & 4844 & 4844 \\ \hline
\textbf{Vulnerabilities} & \textit{VulnerabilitySpan} & code $\Rightarrow$ code       & 2   & 0.4\% & 410  & 410  \\ \bottomrule
\vspace{-0.90cm}
\end{tabular}

}
\end{table}
\subsection{Successful Use Cases}\label{sec:cases}
\textit{\textbf{Galeras}}\cite{galeras}: Galeras is a benchmark for measuring the causal effect of SE prompts for code completion. Galeras configures a set of treatments to assess the influence of potential confounders on the outcomes of ChatGPT (\ie GPT-4). The selected confounders are: \textit{prompt\_size} (from prompts), \textit{n\_whitespaces} (from documentation), \textit{token\_counts}, and \textit{nloc} (from code\_features). This use case of \snipgen demonstrates that prompt engineering strategies (such as those listed in \tabref{tab:templates} - processing prompt) have distinct causal effects on the performance of ChatGPT.

\textit{\textbf{SyntaxEval}}~\cite{syntax_capabilities}: In this use case \textit{Syn taxEval} evaluates the ability of Masked Language Models (\ie Encoder-based Transformers) to predict tokens associated with specific types in the AST representation (\ie syntactic features). \textit{SyntaxEval} used \snipgen to construct a \textbf{\textit{code completion}} testbed with approximately $50K$ Python snippets. \textit{SyntaxEval} aims to account for potential confounders such as \textit{ast\_data} and \textit{code\_features} (illustrated in \figref{fig:diagram}), the analysis revealed no evidence that the evaluated syntactic features influenced the accuracy of the selected models' predictions.

\textit{\textbf{ASTxplainer}}~\cite{astexplainer}: \textit{ASTxplainer} is an explainability method designed to assess how effectively a \llm (\eg decoder-based transformers) predicts syntactic structures. \textit{ASTxplainer} aggregates next-token prediction values through syntactic decomposition, quantified as AsC-Eval values to evaluate the effectiveness. \textit{ASTxplainer} findings reveal that the ability to predict syntactic structures strongly depends on the \llm's parameter size and fine-tuning strategy. Furthermore, causal analysis controlling for confounding variables (e.g., \textit{ast\_data} and \textit{code\_features}) shows that AsC-Eval values at the snippet level negatively impact the cross-entropy loss of the evaluated \llms.

\section{Similar Datasets}\label{sec:similar_datasets}


Significant efforts have produced datasets for evaluating \llms in SE tasks, including DeepFix for program repair \cite{gupta_deepfix_2017}, CodeContest and CoNaLa for program synthesis \cite{li_competition-level_2022, yin2018mining}, and \textit{CodeSearchNet} for code retrieval \cite{husain2019codesearchnet}. Expansions like \textit{CodeXGLUE} \cite{lu_codexglue_2021}, xCodeEval \cite{khan_xcodeeval_2023} target broader tasks, while benchmarks such as \textit{HumanEval} and SecurityEval focus on functional correctness and vulnerabilities \cite{chen_evaluating_2021, secEval}. Despite these efforts, existing datasets often suffer from contamination\cite{jain_livecodebench_2024,yadav_pythonsaga_2024}, with overlaps between training and evaluation data, and benchmarks are prone to memorization by models\cite{ramos2024largelanguagemodelsmemorizing}, limiting their effectiveness in assessing true generalization.

Recent research work has explored the dynamic generation of prompts and testbeds, for instance, \textit{EvoPrompt} is a framework for automatic discrete prompt optimization that connects \llms with Evolutionary Algorithms\cite{guo_connecting_2024}. \textit{Evol-instruct} is a systematic approach to generate instruction-response pairs by iteratively improving prompts and responses through model self-enhancement\cite{xu_wizardlm_2023}. \textit{LiveCodeBench} is a benchmark for evaluating \llms designed to generate code\cite{jain_livecodebench_2024}. Unlike \snipgen, LiveCodeBench addresses issues of data contamination by using continuously updated problems from online coding competitions.

\section{Limitations and future work}\label{sec:limitations}
\textbf{Documentation Quality Analysis:} Meaningfulness evaluation for the \docstring and linked code can not be automatized and depends on the project context. To handle this limitation, we conducted a manual validation. As part of future work, \snipgen should streamline the manual validation process and ascertain the significance of comments and documentation within the snippets.

\textbf{Vulnerability Detection:} The detection of vulnerabilities is reliant solely on the CodeQL tool and its updates; we did not employ any alternative tools to validate these results.

\textbf{Assumption Regarding Snippet Exposure}: \snipgen mitigates to select ``contaminated'' data (\ie already seen snippets) by selecting snippets from specific commit time windows. A practitioner can specify the time windows depending on the \llm release date. \snipgen aims to reduce data contamination by including a prompt and variating the cut code. However, it's important to note that the extracted code changes might include older lines of code or reused code fragments. Our evaluation does not encompass the entire project history to identify older references.

For future work, 1) we propose to extend this dataset to support multiple programming languages; 2) Integrate a wider number of SE tasks. 3) Rank each data point according to the cyclomatic complexity number of AST nodes, documentation, and number of identifiers. The rank will prove better criteria on which snippets are more interesting to evaluate the \llm.

\bibliographystyle{IEEEtran}
\bibliography{utils/citations_bib}

\end{document}